\documentclass[aip,apl,reprint]{revtex4-1}
\usepackage[version=3]{mhchem} 
\usepackage{natbib}
\usepackage{soul}
\usepackage{color}
\usepackage{epstopdf}
\usepackage{graphicx}

\begin{document}
\title{Nucleation and growth of HfO$_{2}$ layers on graphene by CVD}

\author{Grzegorz Lupina}

\author{Mindaugas Lukosius}
\author{Julia Kitzmann}
\author{Jarek Dabrowski}
\author{Andre Wolff}
\author{Wolfgang Mehr}
\affiliation{IHP, Im Technologiepark 25, 15236 Frankfurt (Oder), Germany}

\begin{abstract}

We investigate a seed layer-free growth of HfO$_{2}$ on commercially available chemical vapor deposited (CVD) graphene from various suppliers. It is revealed that the samples of monolayer graphene transferred from Cu to SiO$_{2}$/Si substrates have different coverage with bi- and multi-layer graphene islands. We find that the distribution and number of such islands impact the nucleation and growth of HfO$_{2}$ by CVD. In particular, we show that the edges and surface of densely distributed bi-layer graphene islands provide good nucleation sites for conformal CVD HfO$_{2}$ layers. Dielectric constant of 16 is extracted from measurements on graphene-HfO$_{2}$-TiN capacitors.

\end{abstract}

\maketitle

Many envisioned microelectronic applications of graphene require thin conformal dielectric and semiconductor layers to be grown on possibly highest quality graphene \cite{Grapheneroadmap2012,samsungnature11,reviewferrari2012}. In particular, graphene-based device concepts such as field-effect transistors \cite{IBM2012,Schwierz2013,Sam2013SSE}, vertical field-effect transistors \cite{Brittnell2012}, graphene base transistors \cite{Mehr2012,Sam2012Nanolett,Driussi2013gbt}, and also optoelectronic devices\cite{photonicsreview2012} often involve stacks of materials in which graphene is sandwiched between insulators or semiconductors. Obviously this calls for a method enabling deposition of these materials on graphene, preferrably using methods compatible with the mainstream Si processing. This turns out to be particularly challenging for chemical deposition methods due to the inert nature of graphene surface resulting in nucleation problems\cite{nucleationWang2008}. Although direct growth of dielectrics on graphene by PECVD\cite{IBM2010si3n4} and ALD\cite{noseedHfO2Zou2010} was reported, deposition of conformal insulating films usually requires formation of a seed layer before the actual process of dielectric growth by chemical techniques is started. Various seeding scenarios involving deposition of a polymer layer \cite{Farmer2009polymer,polymerseed2012Shin}, evaporation of a thin metal film\cite{seedmetalKim2009}, or substrate induced seeding\cite{Dlubak2012_Subinduced} were reported so far. The application of seeding significantly improves nucleation behavior, however, the reported methods add complexity and are rarely compatible with standard Si technology.\newline 
In this work, we investigate the nucleation of HfO$_2$ layers on commercially available graphene grown on Cu. Although it requires a transfer, graphene grown on Cu is usually of good quality and provides a useful platform for large area device prototyping. We find that samples from different suppliers transferred to SiO$_2$ substrates, while showing very good crystalline quality (as proved by Raman), differ significantly in surface morphology i.e. the amount and distribution of bi- and multi-layer graphene islands. Furthermore, we demonstrate that not only the edges but also the surface of bi-layer islands found commonly on CVD graphene can provide natural nucleation sites for the CVD growth of HfO$_2$. Our findings implicate that controlling the growth of that kind of islands on the first layer of graphene during CVD process on Cu may provide an effective way to deposit thin conformal layers of various materials on graphene by chemical methods.\newline Graphene on Cu was obtained from three different suppliers. In the following, we will refer to these graphene samples as Graphene A, Graphene B, and Graphene C. About $1 \times 1$cm graphene pieces were transferred from Cu substrates onto 8-inch 300nm SiO$_2$/Si(100) wafers using wet transfer methods described previously\cite{liang2011,TransferRuoff2011}. Briefly, PMMA covered graphene/Cu was placed in ammonium persulfate solution to remove the metal catalyst layer. PMMA-graphene stack was then rinsed in DI water and transferred to the target substrate. This was followed by removal of the polymer layer in acetone bath and IPA-rinse. After transfer, samples were annealed for 30 min at $500^\circ$C in forming gas to clean residual polymer contamination. HfO$_2$ growth was performed in an atomic vapor deposition tool at the substrate temperature of $400^\circ$C with Hf(NMeEt)$_4$ precursor and oxygen as the reactive gas. Raman mapping was done with a Renishaw inVia microscope using $514$ nm laser light and 1800 lines/mm grating. AFM measurements were accomplished using Veeco Digital Instruments Dimension $5000$ in tapping mode at ambient conditions with a Si probe tip. Step heights between $1$-, $2$-, and $3$-layers of graphite exfoliated onto \ce{SiO2}/Si were used to calibrate the height in step measurements. Electrical measurements were performed using Keithley 4200 SCS analyzer.

\begin{figure}[h]
\includegraphics[width=0.7\columnwidth]{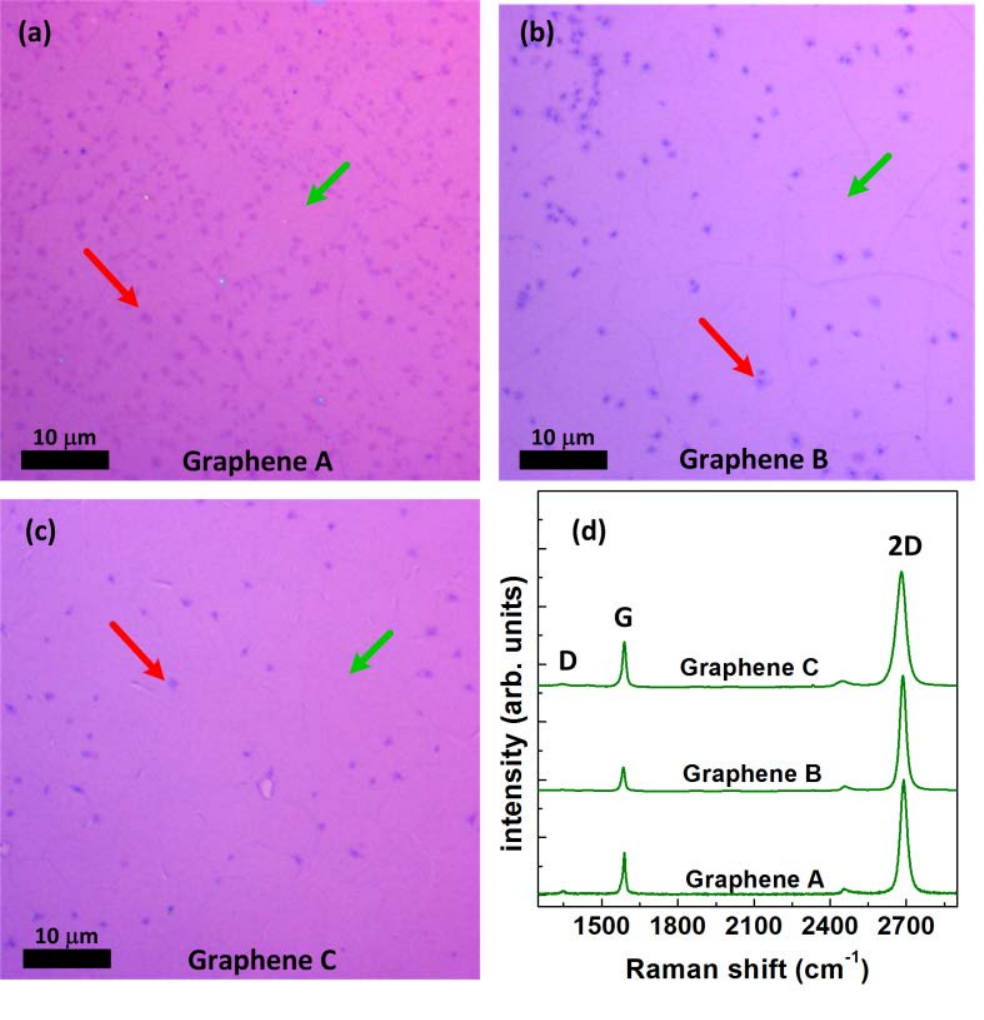}

\vspace{-10pt}

\caption{\label{fig1} Optical microscope images of three different graphene sorts transferred onto 300 nm SiO$_2$/Si substrates (a-c). Red and green arrows indicate examples of multilayer and monolayer regions, respectively. Raman spectra from a monolayer region of each sample (d)}
\end{figure}

Figure 1(a-c) shows the optical microscope images of graphene samples transferred to 300 nm SiO$_2$ / Si substrates. All graphene layers are continous with a small amount of holes and cracks. Each of the samples shows also dark islands with different distributions which are identified as bi- or multi-layer graphene \cite{bilayerRuoff2012,bilayerislandsLiu2012,bilayerislandsYan2011,bilayerislandChung2013}. Graphene A shows the highest density of islands on the underlaying monolayer. Judging from the optical contrast the relatively homogeneously distributed islands are mostly bi-layers. As we will show later, this is also confirmed by AFM. In contrast, on Graphene B the initial monolayer is decorated with multilayer islands (mostly 3-layers with occasionally occuring bi-layers). Clearly, regions with very high and very low density of islands can be distinguished. Graphene C is characterized by a relatively very small areal density of small multilayer islands. The representative Raman spectra measured on the monolayer region after transfer are shown in Fig. 1(d). The presence of strong and narrow G and 2D bands (FWHM(G): 14-20 cm$^{-1}$, FWHM(2D): 32-52 cm$^{-1}$) and a very weak intensity of the defect-related D band confirm high quality of the material. Usually, intensity of the D band increases slightly after forming gas annealing (N$_2$/H$_2$, $500^\circ$C) which is routinely used to remove PMMA residuals before CVD experiments.

\begin{figure}[h]
\includegraphics[width=0.8\columnwidth]{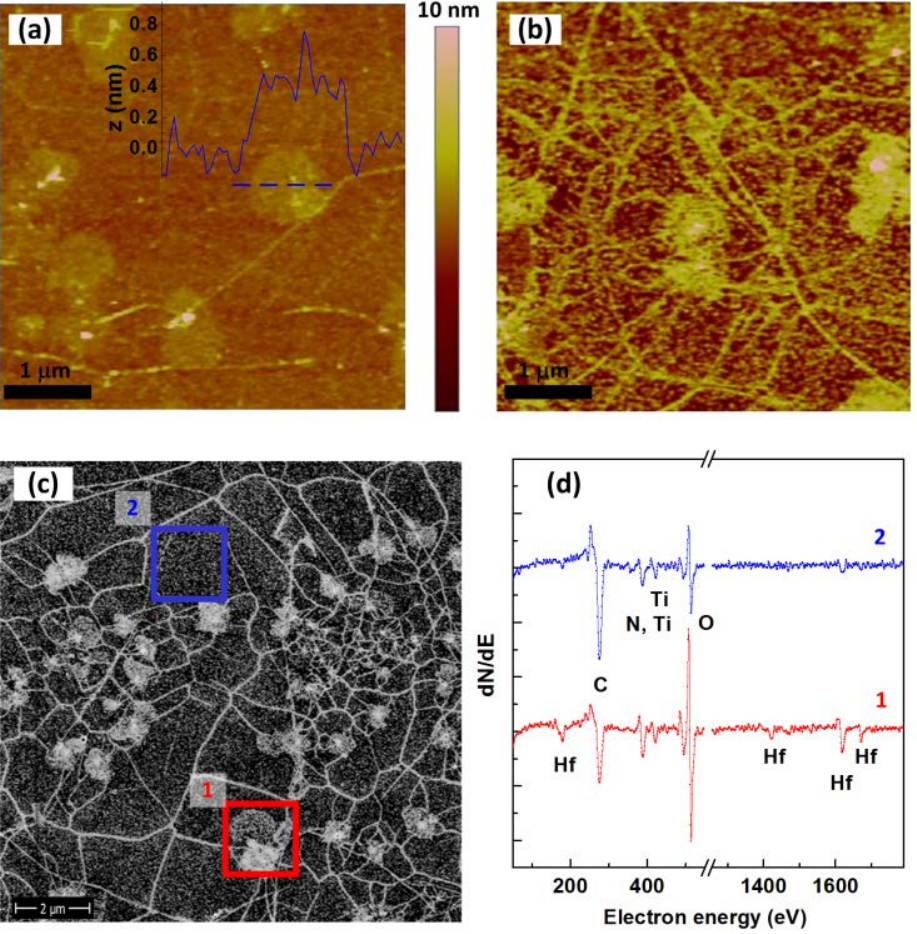}

\vspace{-10pt}

\caption{\label{fig2} AFM image of Graphene A taken before (a) and after (b) deposition of 5 nm HfO$_2$. SEM micrograph of Graphene A after deposition of 5 nm HfO$_2$ (c) and corresponding Auger electron spectra acquired from two characteristic areas 1 and 2 marked with red and blue rectangles (d)}
\end{figure}

Figure 2(a) and (b) shows AFM images of Graphene A acquired before and after deposition of nominally 5 nm of HfO$_{2}$, respectively. Nominal thickness is defined as the thickness of HfO$_{2}$ obtained on a clean SiO$_{2}$ surface neighbouring the graphene flake. Before the deposition (Fig. 2(a)), beside graphene wrinkles also the bi-layer graphene islands can be clearly recognised as elevated rounded areas. The rms roughness between the islands is typically below 0.25nm while on the islands it increases to 0.3 - 0.5nm. Step size measurement was attempted on some of the islands giving usually a very noisy scans. An example of such measurement is shown in the inset to Fig. 2(a) from which a step size accounting for about 1 graphene monolayer can be estimated. After deposition of nominally 5nm of HfO$_{2}$ (Fig. 2(b)), the wrinkles and the surface of the bilayer islands are decorated with white spots which are assigned to HfO$_{2}$. The rms roughness on the monolayer graphene region increases to about 1.1 nm while on the bi-layer islands it is lower with values of about 0.8nm. This indicates that the nucleation of HfO$_{2}$ is more homogeneous on the islands than between them. This is corroborated by SEM and AES results presented in Fig. 2(c) and 2(d), respectively. In the SEM image again two regions are distinguishable: brighter (wrinkles and islands) and darker (monolayer areas between islands and wrinkles). The AES scans (Fig. 2(d)) from bright and dark regions (marked in Fig. 2(c) as area 1 and 2, respectively) show that more intense Hf and O signals along with a stronger attenuated C signal from graphene is detected on bright areas. This confirms that beside graphene wrinkles the HfO$_{2}$ deposit is mainly found on the bilayer islands.

\begin{figure}[h]
\includegraphics[width=0.7\columnwidth]{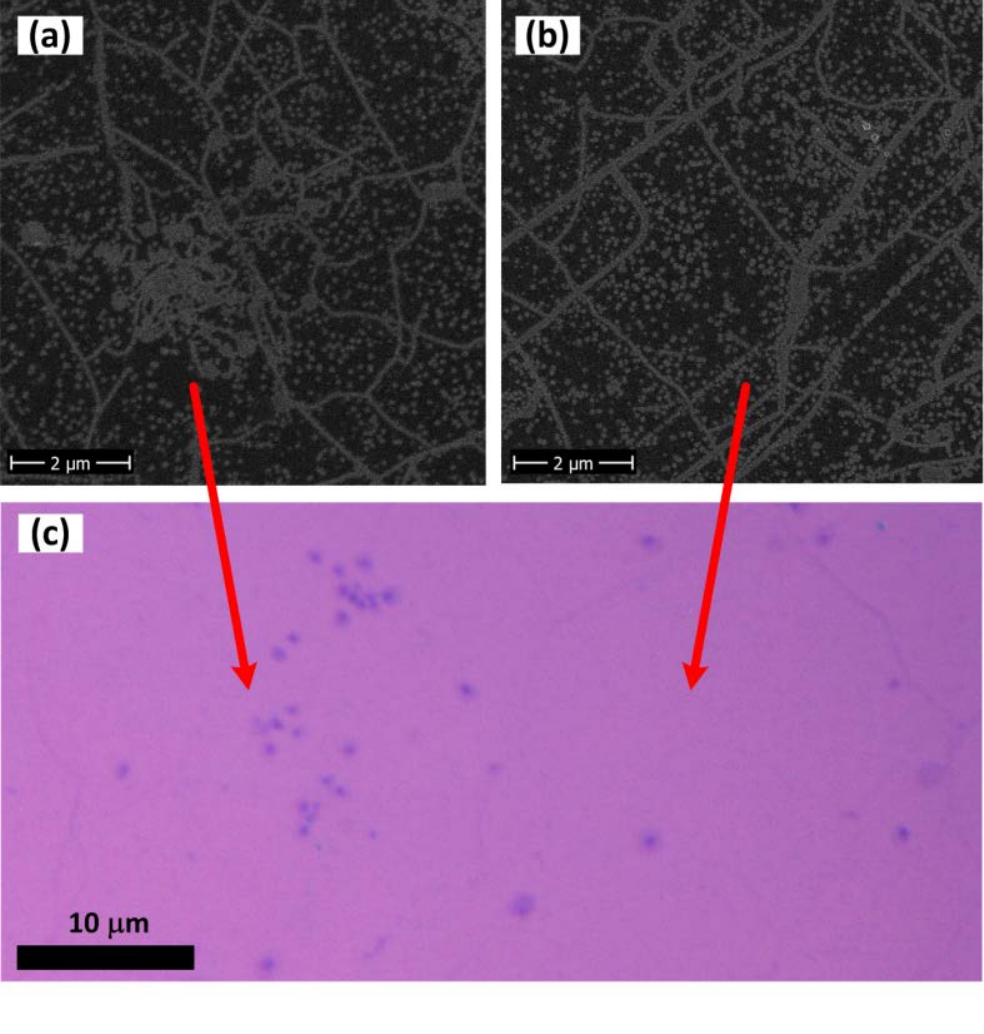}

\vspace{-10pt}

\caption{\label{fig3} Secondary electron micrographs after deposition of 5 nm HfO$_2$ onto Graphene B (a,b), optical microscope image of the initial graphene surface (c). Arrows indicate characteristic regions on the clean graphene to which images in (a) and (b) are correlated}
\end{figure}

Similar behavior is observed for samples of Graphene B (Fig. 3). A difference is that, as stated above, there is a smaller number of bi- and multi-layer islands on Graphene B and that they are distributed in a more inhomogeneous manner. Figure 3(a) and (b) show SEM images taken in two characteristic regions of Graphene B after deposition of 5 nm HfO$_{2}$. In the first region (Fig. 3 (a)), bright oval shapes concentrated in a group can be recognised. In the second region (Fig. 3 b), graphene wrinkles decorated with HfO$_{2}$ are seen, however, the oval features are not present. We correlate these two regions with areas containing high and low number of multilayer islands on the initial graphene surface (cf. Fig. 3(c)). In general, the HfO$_{2}$ layer nucleates better on Graphene A than on Graphene B. This may be due to the higher fraction of clean graphene monolayer (without wrinkles and multilayer islands) which limits the number of nucleation sites in case of Graphene B. Furthermore we find that the more homogeneous distribution of nucleation centers on Graphene A reduces the time after which coalescence of growing HfO$_{2}$ islands is achieved.

\begin{figure}[h]
\includegraphics[width=0.8\columnwidth]{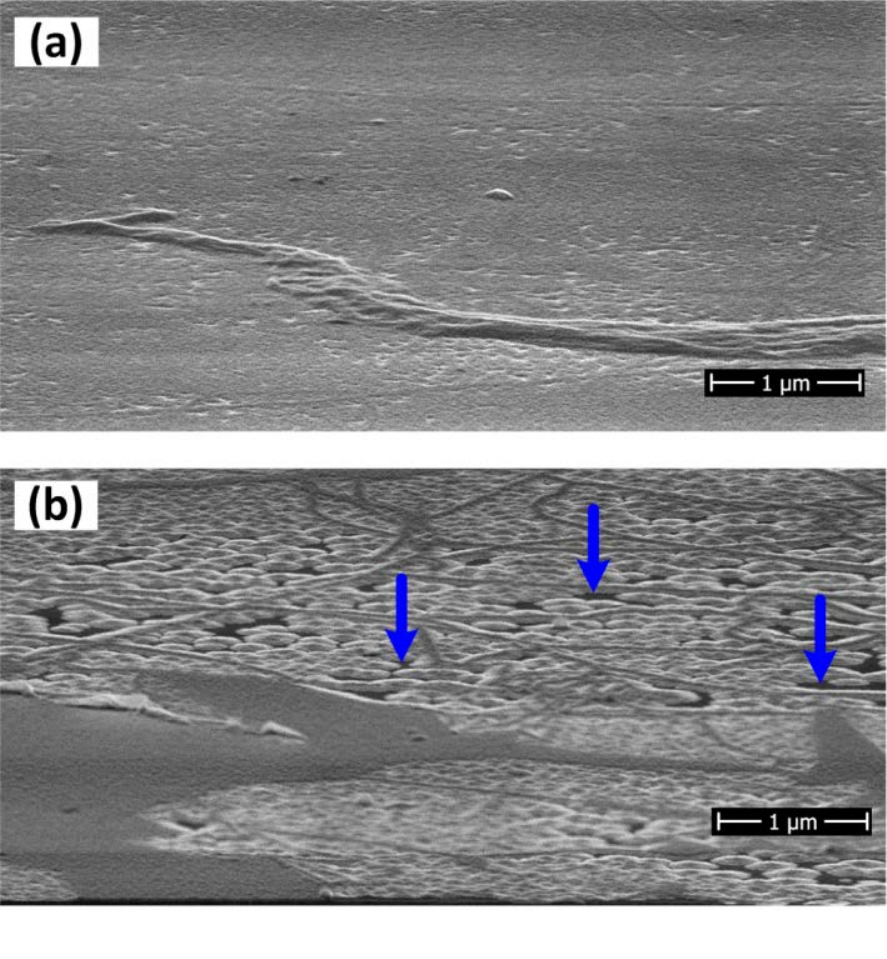}

\vspace{-10pt}

\caption{\label{fig4} SEM images after deposition of 25 nm of HfO$_{2}$ onto Graphene A (a) and Graphene B (b). Blue arrows in (b) indicate examples of areas where the coalescence of HfO$_{2}$ islands is not complete.}
\end{figure}

\noindent This is illustrated in Figure 4 which shows SEM images after deposition of nominally 25 nm of HfO$_{2}$. While HfO$_{2}$ layer on Graphene A is already closed (Fig. 4(a)), full coalescence is not yet achieved on Graphene B (Fig. 4(b)). 
Electrical measurements show that Graphene(A)-HfO$_{2}$-TiN capacitors with 25 nm HfO$_{2}$ are still very poor with high failure rate. It can be a consequence of pinholes existing occasionally in the dielectric layer. For higher HfO$_{2}$ thickness, the yield is greatly improved and good dielectric properties are measured. Figure 5(a) shows an example of capacitance voltage curve measured at 10 kHz for 50 nm HfO$_{2}$ on Graphene A. At this insulator thickness, the total capacitance is dominated by the oxide capacitance and a characteristic modulation due to the quantum capacitance of graphene is not observed\cite{CQXIA2009,CQAppenzeller2008}. Instead, the measured capacitance remains constant with increasing voltage in a way typical for conventional metal-insulator-metal capacitors\cite{Wenger2009}. Figure 5(b) shows that the capacitance scales very well with the capacitor plate area. The dielectric constant extracted from these measurements is 16 which is in good agreement with values reported previously for HfO$_{2}$ on graphene \cite{Meric2008} and metallic substrates\cite{Wenger2008}. 

\begin{figure}[h]
\includegraphics[width=0.95\columnwidth]{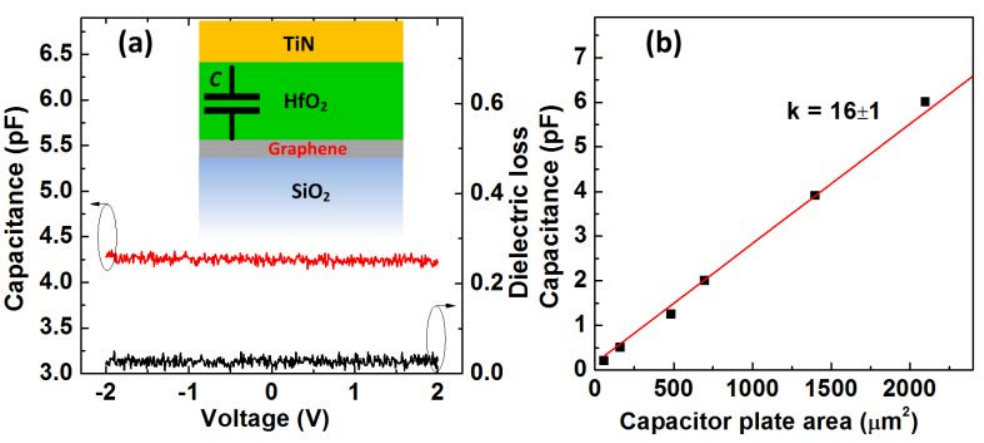}

\vspace{-10pt}

\caption{\label{fig5} Electrical characterization of graphene-HfO$_{2}$-TiN capacitors. Example of capacitance-voltage curve for Graphene A and 50 nm HfO$_{2}$ measured with 10 kHz ac signal (a). Inset in (a) shows a schematic cross-section of the capacitor. Capacitance as a function of the capacitor area (b). Extracted dielectric constant is 16.}
\end{figure}
 
We have shown above that the nucleation of HfO$_{2}$ is significantly better on graphene with high number and homogeneous distribution of bi-layer graphene islands and that thin dielectric layers with good electrical quality can be obtained directly on graphene without any seeding layer. While the exact growth mechanism of CVD dielectrics on such type of graphene is still under investigation our Raman measurements show that the relatively good nucleation on the bi-layer islands can be due to a higher amount of defects present in this region. Such defects can serve as nucleation centers similarly to the nucleation of ALD layers on the edges of graphene flakes as reported before \cite{nucleationWang2008}. Figure 6(a) shows a Raman D-band intensity distribution map taken from a  $20 \times$ 20$\mu$m large area of Graphene A. Figure 6(b) shows the corresponding optical microscope image on which the darker regions represent graphene bi-layer islands as discussed above. There is a very good correlation between the position of the islands and the intensity of the Raman D-band: the intensity of the D-band peaks on the islands and has lowest values on the areas where islands are absent (i.e. in the regions of predominantly monolayer graphene).
Similar nucleation behavior is observed in our CVD experiments focusing on the growth of silicon layers on graphene by using disilane precursor (not shown here). In this case, Si growth proceeds also mainly on the bilayer islands resulting in locally closed and relatively smooth Si layers already in the initial growth stage. In contrast, in the monolayer regions nucleation is more difficult and a growth of separated Si islands takes place. This similarity to HfO$_{2}$ nucleation indicates that our conclusions may be valid for CVD of a broader variety of insulating and semiconducting materials on CVD graphene.

\begin{figure}[h]
\includegraphics[width=0.95\columnwidth]{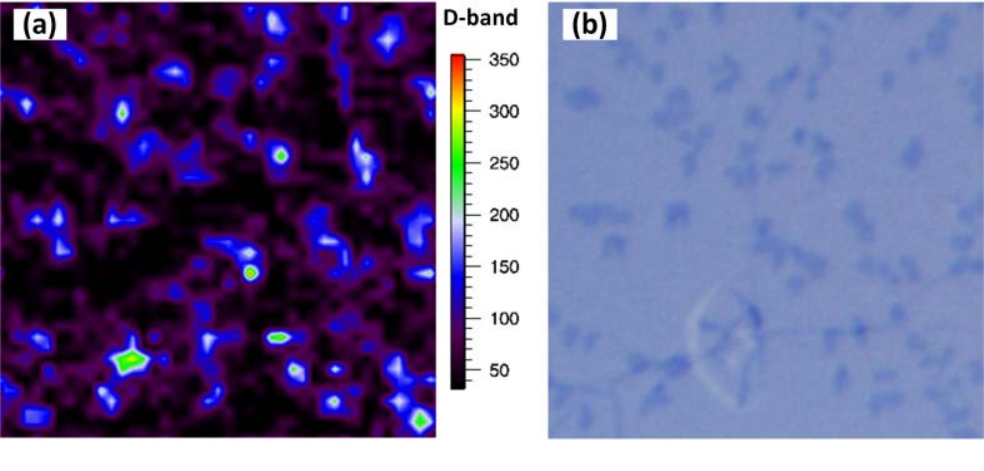}

\vspace{-10pt}

\caption{\label{fig6} Raman D-band intensity map from Graphene A directly before HfO$_{2}$ deposition (a). Analysed area is $20 \times$ 20$\mu$m. Optical microscope image of the same region (b). A very good correlation between an increased intensity of the Raman D-band and the position of the dark islands (bi-layer graphene) in the optical microscope image can be seen.}
\end{figure}

In summary, chemical vapor deposition of HfO$_2$ on commercially available large area graphene  was investigated. We have found that graphene samples from different suppliers transferred to SiO$_2$ substrates show in general a good quality but differ significantly in surface morphology i.e. the amount and distribution of bi- and multi-layer graphene islands. We demonstrated that not only the edges but also the surface of bi-layer islands found commonly on CVD graphene can provide natural nucleation sites for the CVD growth of HfO$_2$. Our findings implicate that controlling the growth of that kind of islands on the first layer of graphene during CVD process on Cu may provide an effective way to deposit thin conformal layers of various materials on graphene by chemical methods and thus reduce the hurdle for the integration of graphene with Si microelectronics. 
 
\vspace{10pt}

The authors thank O. Fursenko, I. Costina, H.-P. Stoll, Y. Yamamoto, and M. Fraschke for experimental support and discussions. Support from the European Commission through a STREP project (GRADE, No. 317839) is gratefully acknowledged.


\begin{thebibliography}{29}%
\makeatletter
\providecommand \@ifxundefined [1]{%
 \@ifx{#1\undefined}
}%
\providecommand \@ifnum [1]{%
 \ifnum #1\expandafter \@firstoftwo
 \else \expandafter \@secondoftwo
 \fi
}%
\providecommand \@ifx [1]{%
 \ifx #1\expandafter \@firstoftwo
 \else \expandafter \@secondoftwo
 \fi
}%
\providecommand \natexlab [1]{#1}%
\providecommand \enquote  [1]{``#1''}%
\providecommand \bibnamefont  [1]{#1}%
\providecommand \bibfnamefont [1]{#1}%
\providecommand \citenamefont [1]{#1}%
\providecommand \href@noop [0]{\@secondoftwo}%
\providecommand \href [0]{\begingroup \@sanitize@url \@href}%
\providecommand \@href[1]{\@@startlink{#1}\@@href}%
\providecommand \@@href[1]{\endgroup#1\@@endlink}%
\providecommand \@sanitize@url [0]{\catcode `\\12\catcode `\$12\catcode
  `\&12\catcode `\#12\catcode `\^12\catcode `\_12\catcode `\%12\relax}%
\providecommand \@@startlink[1]{}%
\providecommand \@@endlink[0]{}%
\providecommand \url  [0]{\begingroup\@sanitize@url \@url }%
\providecommand \@url [1]{\endgroup\@href {#1}{\urlprefix }}%
\providecommand \urlprefix  [0]{URL }%
\providecommand \Eprint [0]{\href }%
\providecommand \doibase [0]{http://dx.doi.org/}%
\providecommand \selectlanguage [0]{\@gobble}%
\providecommand \bibinfo  [0]{\@secondoftwo}%
\providecommand \bibfield  [0]{\@secondoftwo}%
\providecommand \translation [1]{[#1]}%
\providecommand \BibitemOpen [0]{}%
\providecommand \bibitemStop [0]{}%
\providecommand \bibitemNoStop [0]{.\EOS\space}%
\providecommand \EOS [0]{\spacefactor3000\relax}%
\providecommand \BibitemShut  [1]{\csname bibitem#1\endcsname}%
\let\auto@bib@innerbib\@empty
\bibitem [{\citenamefont {Novoselov}\ \emph {et~al.}(2012)\citenamefont
  {Novoselov}, \citenamefont {Falko}, \citenamefont {Colombo}, \citenamefont
  {Gellert}, \citenamefont {Schwab},\ and\ \citenamefont
  {Kim}}]{Grapheneroadmap2012}%
  \BibitemOpen
  \bibfield  {author} {\bibinfo {author} {\bibfnamefont {K.}~\bibnamefont
  {Novoselov}}, \bibinfo {author} {\bibfnamefont {V.}~\bibnamefont {Falko}},
  \bibinfo {author} {\bibfnamefont {L.}~\bibnamefont {Colombo}}, \bibinfo
  {author} {\bibfnamefont {P.}~\bibnamefont {Gellert}}, \bibinfo {author}
  {\bibfnamefont {M.}~\bibnamefont {Schwab}}, \ and\ \bibinfo {author}
  {\bibfnamefont {K.}~\bibnamefont {Kim}},\ }\href@noop {} {\bibfield
  {journal} {\bibinfo  {journal} {Nature}\ }\textbf {\bibinfo {volume} {490}},\
  \bibinfo {pages} {192} (\bibinfo {year} {2012})}\BibitemShut {NoStop}%
\bibitem [{\citenamefont {Kim}\ \emph {et~al.}(2011)\citenamefont {Kim},
  \citenamefont {Choi}, \citenamefont {Kim}, \citenamefont {Cho},\ and\
  \citenamefont {Chung}}]{samsungnature11}%
  \BibitemOpen
  \bibfield  {author} {\bibinfo {author} {\bibfnamefont {K.}~\bibnamefont
  {Kim}}, \bibinfo {author} {\bibfnamefont {J.-Y.}\ \bibnamefont {Choi}},
  \bibinfo {author} {\bibfnamefont {T.}~\bibnamefont {Kim}}, \bibinfo {author}
  {\bibfnamefont {S.-H.}\ \bibnamefont {Cho}}, \ and\ \bibinfo {author}
  {\bibfnamefont {H.-J.}\ \bibnamefont {Chung}},\ }\href@noop {} {\bibfield
  {journal} {\bibinfo  {journal} {Nature}\ }\textbf {\bibinfo {volume} {479}},\
  \bibinfo {pages} {338} (\bibinfo {year} {2011})}\BibitemShut {NoStop}%
\bibitem [{\citenamefont {Bonaccorso}\ \emph {et~al.}(2012)\citenamefont
  {Bonaccorso}, \citenamefont {Lombardo}, \citenamefont {Hasan}, \citenamefont
  {Sun}, \citenamefont {Colombo},\ and\ \citenamefont
  {Ferrari}}]{reviewferrari2012}%
  \BibitemOpen
  \bibfield  {author} {\bibinfo {author} {\bibfnamefont {F.}~\bibnamefont
  {Bonaccorso}}, \bibinfo {author} {\bibfnamefont {A.}~\bibnamefont
  {Lombardo}}, \bibinfo {author} {\bibfnamefont {T.}~\bibnamefont {Hasan}},
  \bibinfo {author} {\bibfnamefont {Z.}~\bibnamefont {Sun}}, \bibinfo {author}
  {\bibfnamefont {L.}~\bibnamefont {Colombo}}, \ and\ \bibinfo {author}
  {\bibfnamefont {A.}~\bibnamefont {Ferrari}},\ }\href@noop {} {\bibfield
  {journal} {\bibinfo  {journal} {Materials Today}\ }\textbf {\bibinfo {volume}
  {15}},\ \bibinfo {pages} {564} (\bibinfo {year} {2012})}\BibitemShut
  {NoStop}%
\bibitem [{\citenamefont {Wu}\ \emph {et~al.}(2012)\citenamefont {Wu},
  \citenamefont {Jenkins}, \citenamefont {Valdes-Garcia}, \citenamefont
  {Farmer}, \citenamefont {Zhu}, \citenamefont {Bol}, \citenamefont
  {Dimitrakopoulos}, \citenamefont {Zhu}, \citenamefont {Xia}, \citenamefont
  {Avouris},\ and\ \citenamefont {Lin}}]{IBM2012}%
  \BibitemOpen
  \bibfield  {author} {\bibinfo {author} {\bibfnamefont {Y.}~\bibnamefont
  {Wu}}, \bibinfo {author} {\bibfnamefont {K.}~\bibnamefont {Jenkins}},
  \bibinfo {author} {\bibfnamefont {A.}~\bibnamefont {Valdes-Garcia}}, \bibinfo
  {author} {\bibfnamefont {D.}~\bibnamefont {Farmer}}, \bibinfo {author}
  {\bibfnamefont {Y.}~\bibnamefont {Zhu}}, \bibinfo {author} {\bibfnamefont
  {A.}~\bibnamefont {Bol}}, \bibinfo {author} {\bibfnamefont {C.}~\bibnamefont
  {Dimitrakopoulos}}, \bibinfo {author} {\bibfnamefont {W.}~\bibnamefont
  {Zhu}}, \bibinfo {author} {\bibfnamefont {F.}~\bibnamefont {Xia}}, \bibinfo
  {author} {\bibfnamefont {P.}~\bibnamefont {Avouris}}, \ and\ \bibinfo
  {author} {\bibfnamefont {Y.-M.}\ \bibnamefont {Lin}},\ }\href@noop {}
  {\bibfield  {journal} {\bibinfo  {journal} {Nano Letters}\ }\textbf {\bibinfo
  {volume} {12}},\ \bibinfo {pages} {3062} (\bibinfo {year}
  {2012})}\BibitemShut {NoStop}%
\bibitem [{\citenamefont {Schwierz}(2013)}]{Schwierz2013}%
  \BibitemOpen
  \bibfield  {author} {\bibinfo {author} {\bibfnamefont {F.}~\bibnamefont
  {Schwierz}},\ }\href@noop {} {\bibfield  {journal} {\bibinfo  {journal}
  {Proceedings of the IEEE}\ }\textbf {\bibinfo {volume} {101}},\ \bibinfo
  {pages} {1567} (\bibinfo {year} {2013})}\BibitemShut {NoStop}%
\bibitem [{\citenamefont {Vaziri}\ \emph
  {et~al.}(2013{\natexlab{a}})\citenamefont {Vaziri}, \citenamefont {Lupina},
  \citenamefont {Paussa}, \citenamefont {Smith}, \citenamefont {Henkel},
  \citenamefont {Lippert}, \citenamefont {Dabrowski}, \citenamefont {Mehr},
  \citenamefont {Ostling},\ and\ \citenamefont {Lemme}}]{Sam2013SSE}%
  \BibitemOpen
  \bibfield  {author} {\bibinfo {author} {\bibfnamefont {S.}~\bibnamefont
  {Vaziri}}, \bibinfo {author} {\bibfnamefont {G.}~\bibnamefont {Lupina}},
  \bibinfo {author} {\bibfnamefont {A.}~\bibnamefont {Paussa}}, \bibinfo
  {author} {\bibfnamefont {A.}~\bibnamefont {Smith}}, \bibinfo {author}
  {\bibfnamefont {C.}~\bibnamefont {Henkel}}, \bibinfo {author} {\bibfnamefont
  {G.}~\bibnamefont {Lippert}}, \bibinfo {author} {\bibfnamefont
  {J.}~\bibnamefont {Dabrowski}}, \bibinfo {author} {\bibfnamefont
  {W.}~\bibnamefont {Mehr}}, \bibinfo {author} {\bibfnamefont {M.}~\bibnamefont
  {Ostling}}, \ and\ \bibinfo {author} {\bibfnamefont {M.}~\bibnamefont
  {Lemme}},\ }\href@noop {} {\bibfield  {journal} {\bibinfo  {journal} {Solid
  State Electronics}\ }\textbf {\bibinfo {volume} {84}},\ \bibinfo {pages}
  {185} (\bibinfo {year} {2013}{\natexlab{a}})}\BibitemShut {NoStop}%
\bibitem [{\citenamefont {Britnell}\ \emph {et~al.}(2012)\citenamefont
  {Britnell}, \citenamefont {Gorbachev}, \citenamefont {Jalil}, \citenamefont
  {Belle}, \citenamefont {Schedin}, \citenamefont {Mishchenko}, \citenamefont
  {Georgiou}, \citenamefont {Katsnelson}, \citenamefont {Eaves}, \citenamefont
  {Morozov}, \citenamefont {Peres}, \citenamefont {Leist}, \citenamefont
  {Geim}, \citenamefont {Novoselov},\ and\ \citenamefont
  {Ponomarenko}}]{Brittnell2012}%
  \BibitemOpen
  \bibfield  {author} {\bibinfo {author} {\bibfnamefont {L.}~\bibnamefont
  {Britnell}}, \bibinfo {author} {\bibfnamefont {R.~V.}\ \bibnamefont
  {Gorbachev}}, \bibinfo {author} {\bibfnamefont {R.}~\bibnamefont {Jalil}},
  \bibinfo {author} {\bibfnamefont {B.~D.}\ \bibnamefont {Belle}}, \bibinfo
  {author} {\bibfnamefont {F.}~\bibnamefont {Schedin}}, \bibinfo {author}
  {\bibfnamefont {A.}~\bibnamefont {Mishchenko}}, \bibinfo {author}
  {\bibfnamefont {T.}~\bibnamefont {Georgiou}}, \bibinfo {author}
  {\bibfnamefont {M.~I.}\ \bibnamefont {Katsnelson}}, \bibinfo {author}
  {\bibfnamefont {L.}~\bibnamefont {Eaves}}, \bibinfo {author} {\bibfnamefont
  {S.~V.}\ \bibnamefont {Morozov}}, \bibinfo {author} {\bibfnamefont
  {N.~M.~R.}\ \bibnamefont {Peres}}, \bibinfo {author} {\bibfnamefont
  {J.}~\bibnamefont {Leist}}, \bibinfo {author} {\bibfnamefont {A.~K.}\
  \bibnamefont {Geim}}, \bibinfo {author} {\bibfnamefont {K.~S.}\ \bibnamefont
  {Novoselov}}, \ and\ \bibinfo {author} {\bibfnamefont {L.~A.}\ \bibnamefont
  {Ponomarenko}},\ }\href@noop {} {\bibfield  {journal} {\bibinfo  {journal}
  {Science}\ }\textbf {\bibinfo {volume} {335}},\ \bibinfo {pages} {947}
  (\bibinfo {year} {2012})}\BibitemShut {NoStop}%
\bibitem [{\citenamefont {Mehr}\ \emph {et~al.}(2012)\citenamefont {Mehr},
  \citenamefont {Dabrowski}, \citenamefont {Scheytt}, \citenamefont {Lippert},
  \citenamefont {Xie}, \citenamefont {Lemme}, \citenamefont {Ostling},\ and\
  \citenamefont {Lupina}}]{Mehr2012}%
  \BibitemOpen
  \bibfield  {author} {\bibinfo {author} {\bibfnamefont {W.}~\bibnamefont
  {Mehr}}, \bibinfo {author} {\bibfnamefont {J.}~\bibnamefont {Dabrowski}},
  \bibinfo {author} {\bibfnamefont {J.}~\bibnamefont {Scheytt}}, \bibinfo
  {author} {\bibfnamefont {G.}~\bibnamefont {Lippert}}, \bibinfo {author}
  {\bibfnamefont {Y.}~\bibnamefont {Xie}}, \bibinfo {author} {\bibfnamefont
  {M.}~\bibnamefont {Lemme}}, \bibinfo {author} {\bibfnamefont
  {M.}~\bibnamefont {Ostling}}, \ and\ \bibinfo {author} {\bibfnamefont
  {G.}~\bibnamefont {Lupina}},\ }\href@noop {} {\bibfield  {journal} {\bibinfo
  {journal} {Electron Device Letters}\ }\textbf {\bibinfo {volume} {33}},\
  \bibinfo {pages} {691} (\bibinfo {year} {2012})}\BibitemShut {NoStop}%
\bibitem [{\citenamefont {Vaziri}\ \emph
  {et~al.}(2013{\natexlab{b}})\citenamefont {Vaziri}, \citenamefont {Lupina},
  \citenamefont {Henkel}, \citenamefont {Smith}, \citenamefont {Ostling},
  \citenamefont {Dabrowski}, \citenamefont {Lippert}, \citenamefont {Mehr},\
  and\ \citenamefont {Lemme}}]{Sam2012Nanolett}%
  \BibitemOpen
  \bibfield  {author} {\bibinfo {author} {\bibfnamefont {S.}~\bibnamefont
  {Vaziri}}, \bibinfo {author} {\bibfnamefont {G.}~\bibnamefont {Lupina}},
  \bibinfo {author} {\bibfnamefont {C.}~\bibnamefont {Henkel}}, \bibinfo
  {author} {\bibfnamefont {A.}~\bibnamefont {Smith}}, \bibinfo {author}
  {\bibfnamefont {M.}~\bibnamefont {Ostling}}, \bibinfo {author} {\bibfnamefont
  {J.}~\bibnamefont {Dabrowski}}, \bibinfo {author} {\bibfnamefont
  {G.}~\bibnamefont {Lippert}}, \bibinfo {author} {\bibfnamefont
  {W.}~\bibnamefont {Mehr}}, \ and\ \bibinfo {author} {\bibfnamefont
  {M.}~\bibnamefont {Lemme}},\ }\href@noop {} {\bibfield  {journal} {\bibinfo
  {journal} {Nano Letters}\ }\textbf {\bibinfo {volume} {13}},\ \bibinfo
  {pages} {1435} (\bibinfo {year} {2013}{\natexlab{b}})}\BibitemShut {NoStop}%
\bibitem [{\citenamefont {Driussi}, \citenamefont {Palestri},\ and\
  \citenamefont {Selmi}(2013)}]{Driussi2013gbt}%
  \BibitemOpen
  \bibfield  {author} {\bibinfo {author} {\bibfnamefont {F.}~\bibnamefont
  {Driussi}}, \bibinfo {author} {\bibfnamefont {P.}~\bibnamefont {Palestri}}, \
  and\ \bibinfo {author} {\bibfnamefont {L.}~\bibnamefont {Selmi}},\
  }\href@noop {} {\bibfield  {journal} {\bibinfo  {journal} {Microelectronic
  Engineering}\ }\textbf {\bibinfo {volume} {109}},\ \bibinfo {pages} {338}
  (\bibinfo {year} {2013})}\BibitemShut {NoStop}%
\bibitem [{\citenamefont {Bao}\ and\ \citenamefont
  {Loh}(2012)}]{photonicsreview2012}%
  \BibitemOpen
  \bibfield  {author} {\bibinfo {author} {\bibfnamefont {Q.}~\bibnamefont
  {Bao}}\ and\ \bibinfo {author} {\bibfnamefont {K.}~\bibnamefont {Loh}},\
  }\href@noop {} {\bibfield  {journal} {\bibinfo  {journal} {ACS Nano}\
  }\textbf {\bibinfo {volume} {6}},\ \bibinfo {pages} {3677} (\bibinfo {year}
  {2012})}\BibitemShut {NoStop}%
\bibitem [{\citenamefont {Wang}, \citenamefont {Tabakman},\ and\ \citenamefont
  {Dai}(2008)}]{nucleationWang2008}%
  \BibitemOpen
  \bibfield  {author} {\bibinfo {author} {\bibfnamefont {X.}~\bibnamefont
  {Wang}}, \bibinfo {author} {\bibfnamefont {S.}~\bibnamefont {Tabakman}}, \
  and\ \bibinfo {author} {\bibfnamefont {H.}~\bibnamefont {Dai}},\ }\href@noop
  {} {\bibfield  {journal} {\bibinfo  {journal} {J. Am. Chem. Soc.}\ }\textbf
  {\bibinfo {volume} {130}},\ \bibinfo {pages} {8152} (\bibinfo {year}
  {2008})}\BibitemShut {NoStop}%
\bibitem [{\citenamefont {Zhu}\ \emph {et~al.}(2010)\citenamefont {Zhu},
  \citenamefont {Neumayer}, \citenamefont {Perebeinos},\ and\ \citenamefont
  {Avouris}}]{IBM2010si3n4}%
  \BibitemOpen
  \bibfield  {author} {\bibinfo {author} {\bibfnamefont {W.}~\bibnamefont
  {Zhu}}, \bibinfo {author} {\bibfnamefont {D.}~\bibnamefont {Neumayer}},
  \bibinfo {author} {\bibfnamefont {V.}~\bibnamefont {Perebeinos}}, \ and\
  \bibinfo {author} {\bibfnamefont {P.}~\bibnamefont {Avouris}},\ }\href@noop
  {} {\bibfield  {journal} {\bibinfo  {journal} {Nano Letters}\ }\textbf
  {\bibinfo {volume} {10}},\ \bibinfo {pages} {3572} (\bibinfo {year}
  {2010})}\BibitemShut {NoStop}%
\bibitem [{\citenamefont {Zou}\ \emph {et~al.}(2010)\citenamefont {Zou},
  \citenamefont {Hong}, \citenamefont {Keefer},\ and\ \citenamefont
  {Zhu}}]{noseedHfO2Zou2010}%
  \BibitemOpen
  \bibfield  {author} {\bibinfo {author} {\bibfnamefont {K.}~\bibnamefont
  {Zou}}, \bibinfo {author} {\bibfnamefont {X.}~\bibnamefont {Hong}}, \bibinfo
  {author} {\bibfnamefont {D.}~\bibnamefont {Keefer}}, \ and\ \bibinfo {author}
  {\bibfnamefont {J.}~\bibnamefont {Zhu}},\ }\href@noop {} {\bibfield
  {journal} {\bibinfo  {journal} {Physical Review Letters}\ }\textbf {\bibinfo
  {volume} {105}},\ \bibinfo {pages} {126601} (\bibinfo {year}
  {2010})}\BibitemShut {NoStop}%
\bibitem [{\citenamefont {Farmer}\ \emph {et~al.}(2009)\citenamefont {Farmer},
  \citenamefont {Chiu}, \citenamefont {Lin}, \citenamefont {Jenkins},
  \citenamefont {Xia},\ and\ \citenamefont {Avouris}}]{Farmer2009polymer}%
  \BibitemOpen
  \bibfield  {author} {\bibinfo {author} {\bibfnamefont {D.}~\bibnamefont
  {Farmer}}, \bibinfo {author} {\bibfnamefont {H.-Y.}\ \bibnamefont {Chiu}},
  \bibinfo {author} {\bibfnamefont {Y.}~\bibnamefont {Lin}}, \bibinfo {author}
  {\bibfnamefont {K.}~\bibnamefont {Jenkins}}, \bibinfo {author} {\bibfnamefont
  {F.}~\bibnamefont {Xia}}, \ and\ \bibinfo {author} {\bibfnamefont
  {P.}~\bibnamefont {Avouris}},\ }\href@noop {} {\bibfield  {journal} {\bibinfo
   {journal} {Nano Letters}\ }\textbf {\bibinfo {volume} {9}},\ \bibinfo
  {pages} {4474} (\bibinfo {year} {2009})}\BibitemShut {NoStop}%
\bibitem [{\citenamefont {Shin}\ \emph {et~al.}(2012)\citenamefont {Shin},
  \citenamefont {Kim}, \citenamefont {Sul},\ and\ \citenamefont
  {Cho}}]{polymerseed2012Shin}%
  \BibitemOpen
  \bibfield  {author} {\bibinfo {author} {\bibfnamefont {W.~C.}\ \bibnamefont
  {Shin}}, \bibinfo {author} {\bibfnamefont {T.}~\bibnamefont {Kim}}, \bibinfo
  {author} {\bibfnamefont {O.}~\bibnamefont {Sul}}, \ and\ \bibinfo {author}
  {\bibfnamefont {B.}~\bibnamefont {Cho}},\ }\href@noop {} {\bibfield
  {journal} {\bibinfo  {journal} {Applied Physics Letters}\ }\textbf {\bibinfo
  {volume} {101}},\ \bibinfo {pages} {033507} (\bibinfo {year}
  {2012})}\BibitemShut {NoStop}%
\bibitem [{\citenamefont {Kim}\ \emph {et~al.}(2009)\citenamefont {Kim},
  \citenamefont {Nah}, \citenamefont {Jo}, \citenamefont {Shahrjerdi},
  \citenamefont {Colombo}, \citenamefont {Yao}, \citenamefont {Tutuc},\ and\
  \citenamefont {Banerjee}}]{seedmetalKim2009}%
  \BibitemOpen
  \bibfield  {author} {\bibinfo {author} {\bibfnamefont {S.}~\bibnamefont
  {Kim}}, \bibinfo {author} {\bibfnamefont {J.}~\bibnamefont {Nah}}, \bibinfo
  {author} {\bibfnamefont {I.}~\bibnamefont {Jo}}, \bibinfo {author}
  {\bibfnamefont {D.}~\bibnamefont {Shahrjerdi}}, \bibinfo {author}
  {\bibfnamefont {L.}~\bibnamefont {Colombo}}, \bibinfo {author} {\bibfnamefont
  {Z.}~\bibnamefont {Yao}}, \bibinfo {author} {\bibfnamefont {E.}~\bibnamefont
  {Tutuc}}, \ and\ \bibinfo {author} {\bibfnamefont {S.}~\bibnamefont
  {Banerjee}},\ }\href@noop {} {\bibfield  {journal} {\bibinfo  {journal}
  {Applied Physics Letters}\ }\textbf {\bibinfo {volume} {94}},\ \bibinfo
  {pages} {062107} (\bibinfo {year} {2009})}\BibitemShut {NoStop}%
\bibitem [{\citenamefont {Dlubak}\ \emph {et~al.}(2012)\citenamefont {Dlubak},
  \citenamefont {Kidambi}, \citenamefont {Weatherup}, \citenamefont {Hofmann},\
  and\ \citenamefont {Robertson}}]{Dlubak2012_Subinduced}%
  \BibitemOpen
  \bibfield  {author} {\bibinfo {author} {\bibfnamefont {B.}~\bibnamefont
  {Dlubak}}, \bibinfo {author} {\bibfnamefont {P.}~\bibnamefont {Kidambi}},
  \bibinfo {author} {\bibfnamefont {R.}~\bibnamefont {Weatherup}}, \bibinfo
  {author} {\bibfnamefont {S.}~\bibnamefont {Hofmann}}, \ and\ \bibinfo
  {author} {\bibfnamefont {J.}~\bibnamefont {Robertson}},\ }\href@noop {}
  {\bibfield  {journal} {\bibinfo  {journal} {Applied Physics Letters}\
  }\textbf {\bibinfo {volume} {100}},\ \bibinfo {pages} {173113} (\bibinfo
  {year} {2012})}\BibitemShut {NoStop}%
\bibitem [{\citenamefont {Liang}\ \emph {et~al.}(2011)\citenamefont {Liang},
  \citenamefont {Sperling}, \citenamefont {Calizo}, \citenamefont {Cheng},
  \citenamefont {Hacker}, \citenamefont {Zhang}, \citenamefont {Obeng},
  \citenamefont {Yan}, \citenamefont {Peng}, \citenamefont {Li}, \citenamefont
  {Zhu}, \citenamefont {Yuan}, \citenamefont {Walker}, \citenamefont {Liu},
  \citenamefont {Peng},\ and\ \citenamefont {Richter}}]{liang2011}%
  \BibitemOpen
  \bibfield  {author} {\bibinfo {author} {\bibfnamefont {X.}~\bibnamefont
  {Liang}}, \bibinfo {author} {\bibfnamefont {B.~A.}\ \bibnamefont {Sperling}},
  \bibinfo {author} {\bibfnamefont {I.}~\bibnamefont {Calizo}}, \bibinfo
  {author} {\bibfnamefont {G.}~\bibnamefont {Cheng}}, \bibinfo {author}
  {\bibfnamefont {C.~A.}\ \bibnamefont {Hacker}}, \bibinfo {author}
  {\bibfnamefont {Q.}~\bibnamefont {Zhang}}, \bibinfo {author} {\bibfnamefont
  {Y.}~\bibnamefont {Obeng}}, \bibinfo {author} {\bibfnamefont
  {K.}~\bibnamefont {Yan}}, \bibinfo {author} {\bibfnamefont {H.}~\bibnamefont
  {Peng}}, \bibinfo {author} {\bibfnamefont {Q.}~\bibnamefont {Li}}, \bibinfo
  {author} {\bibfnamefont {X.}~\bibnamefont {Zhu}}, \bibinfo {author}
  {\bibfnamefont {H.}~\bibnamefont {Yuan}}, \bibinfo {author} {\bibfnamefont
  {A.~R.~H.}\ \bibnamefont {Walker}}, \bibinfo {author} {\bibfnamefont
  {Z.}~\bibnamefont {Liu}}, \bibinfo {author} {\bibfnamefont {L.-M.}\
  \bibnamefont {Peng}}, \ and\ \bibinfo {author} {\bibfnamefont {C.~A.}\
  \bibnamefont {Richter}},\ }\href@noop {} {\bibfield  {journal} {\bibinfo
  {journal} {ACS Nano}\ }\textbf {\bibinfo {volume} {5}},\ \bibinfo {pages}
  {9144} (\bibinfo {year} {2011})}\BibitemShut {NoStop}%
\bibitem [{\citenamefont {Suk}\ \emph {et~al.}(2011)\citenamefont {Suk},
  \citenamefont {Kitt}, \citenamefont {Magnuson}, \citenamefont {Hao},
  \citenamefont {Ahmed}, \citenamefont {An}, \citenamefont {Swan},
  \citenamefont {Goldberg},\ and\ \citenamefont {Ruoff}}]{TransferRuoff2011}%
  \BibitemOpen
  \bibfield  {author} {\bibinfo {author} {\bibfnamefont {J.}~\bibnamefont
  {Suk}}, \bibinfo {author} {\bibfnamefont {A.}~\bibnamefont {Kitt}}, \bibinfo
  {author} {\bibfnamefont {C.}~\bibnamefont {Magnuson}}, \bibinfo {author}
  {\bibfnamefont {Y.}~\bibnamefont {Hao}}, \bibinfo {author} {\bibfnamefont
  {S.}~\bibnamefont {Ahmed}}, \bibinfo {author} {\bibfnamefont
  {J.}~\bibnamefont {An}}, \bibinfo {author} {\bibfnamefont {A.}~\bibnamefont
  {Swan}}, \bibinfo {author} {\bibfnamefont {B.}~\bibnamefont {Goldberg}}, \
  and\ \bibinfo {author} {\bibfnamefont {R.}~\bibnamefont {Ruoff}},\
  }\href@noop {} {\bibfield  {journal} {\bibinfo  {journal} {ACS Nano}\
  }\textbf {\bibinfo {volume} {5}},\ \bibinfo {pages} {6916} (\bibinfo {year}
  {2011})}\BibitemShut {NoStop}%
\bibitem [{\citenamefont {Kholmanov}\ \emph {et~al.}(2012)\citenamefont
  {Kholmanov}, \citenamefont {Magnuson}, \citenamefont {Aliev}, \citenamefont
  {Li}, \citenamefont {Zhang}, \citenamefont {Suk}, \citenamefont {Zhang},
  \citenamefont {Peng}, \citenamefont {S.~Hossein~Mousavi}, \citenamefont
  {Piner}, \citenamefont {Shvets},\ and\ \citenamefont
  {Ruoff}}]{bilayerRuoff2012}%
  \BibitemOpen
  \bibfield  {author} {\bibinfo {author} {\bibfnamefont {I.}~\bibnamefont
  {Kholmanov}}, \bibinfo {author} {\bibfnamefont {C.}~\bibnamefont {Magnuson}},
  \bibinfo {author} {\bibfnamefont {A.}~\bibnamefont {Aliev}}, \bibinfo
  {author} {\bibfnamefont {H.}~\bibnamefont {Li}}, \bibinfo {author}
  {\bibfnamefont {B.}~\bibnamefont {Zhang}}, \bibinfo {author} {\bibfnamefont
  {J.}~\bibnamefont {Suk}}, \bibinfo {author} {\bibfnamefont {L.}~\bibnamefont
  {Zhang}}, \bibinfo {author} {\bibfnamefont {E.}~\bibnamefont {Peng}},
  \bibinfo {author} {\bibfnamefont {A.~K.}\ \bibnamefont {S.~Hossein~Mousavi}},
  \bibinfo {author} {\bibfnamefont {R.}~\bibnamefont {Piner}}, \bibinfo
  {author} {\bibfnamefont {G.}~\bibnamefont {Shvets}}, \ and\ \bibinfo {author}
  {\bibfnamefont {R.}~\bibnamefont {Ruoff}},\ }\href@noop {} {\bibfield
  {journal} {\bibinfo  {journal} {Nano Letters}\ }\textbf {\bibinfo {volume}
  {12}},\ \bibinfo {pages} {5679} (\bibinfo {year} {2012})}\BibitemShut
  {NoStop}%
\bibitem [{\citenamefont {Liu}\ \emph {et~al.}(2012)\citenamefont {Liu},
  \citenamefont {Zhou}, \citenamefont {Cheng}, \citenamefont {Yu},
  \citenamefont {Liu}, \citenamefont {Chen}, \citenamefont {Shaw},
  \citenamefont {Zhong}, \citenamefont {Huang},\ and\ \citenamefont
  {Duan}}]{bilayerislandsLiu2012}%
  \BibitemOpen
  \bibfield  {author} {\bibinfo {author} {\bibfnamefont {L.}~\bibnamefont
  {Liu}}, \bibinfo {author} {\bibfnamefont {H.}~\bibnamefont {Zhou}}, \bibinfo
  {author} {\bibfnamefont {R.}~\bibnamefont {Cheng}}, \bibinfo {author}
  {\bibfnamefont {W.}~\bibnamefont {Yu}}, \bibinfo {author} {\bibfnamefont
  {Y.}~\bibnamefont {Liu}}, \bibinfo {author} {\bibfnamefont {Y.}~\bibnamefont
  {Chen}}, \bibinfo {author} {\bibfnamefont {J.}~\bibnamefont {Shaw}}, \bibinfo
  {author} {\bibfnamefont {X.}~\bibnamefont {Zhong}}, \bibinfo {author}
  {\bibfnamefont {Y.}~\bibnamefont {Huang}}, \ and\ \bibinfo {author}
  {\bibfnamefont {X.}~\bibnamefont {Duan}},\ }\href@noop {} {\bibfield
  {journal} {\bibinfo  {journal} {ACS Nano}\ }\textbf {\bibinfo {volume} {6}},\
  \bibinfo {pages} {8241} (\bibinfo {year} {2012})}\BibitemShut {NoStop}%
\bibitem [{\citenamefont {Yan}\ \emph {et~al.}(2011)\citenamefont {Yan},
  \citenamefont {Peng}, \citenamefont {Zhou}, \citenamefont {Li},\ and\
  \citenamefont {Liu}}]{bilayerislandsYan2011}%
  \BibitemOpen
  \bibfield  {author} {\bibinfo {author} {\bibfnamefont {K.}~\bibnamefont
  {Yan}}, \bibinfo {author} {\bibfnamefont {H.}~\bibnamefont {Peng}}, \bibinfo
  {author} {\bibfnamefont {Y.}~\bibnamefont {Zhou}}, \bibinfo {author}
  {\bibfnamefont {H.}~\bibnamefont {Li}}, \ and\ \bibinfo {author}
  {\bibfnamefont {Z.}~\bibnamefont {Liu}},\ }\href@noop {} {\bibfield
  {journal} {\bibinfo  {journal} {Nano Letters}\ }\textbf {\bibinfo {volume}
  {11}},\ \bibinfo {pages} {1106} (\bibinfo {year} {2011})}\BibitemShut
  {NoStop}%
\bibitem [{\citenamefont {Chung}\ \emph {et~al.}(2013)\citenamefont {Chung},
  \citenamefont {Shen}, \citenamefont {Cao}, \citenamefont {Jauregui},
  \citenamefont {Wu}, \citenamefont {Yu}, \citenamefont {Newell},\ and\
  \citenamefont {Chen}}]{bilayerislandChung2013}%
  \BibitemOpen
  \bibfield  {author} {\bibinfo {author} {\bibfnamefont {T.}~\bibnamefont
  {Chung}}, \bibinfo {author} {\bibfnamefont {T.}~\bibnamefont {Shen}},
  \bibinfo {author} {\bibfnamefont {H.}~\bibnamefont {Cao}}, \bibinfo {author}
  {\bibfnamefont {L.}~\bibnamefont {Jauregui}}, \bibinfo {author}
  {\bibfnamefont {W.}~\bibnamefont {Wu}}, \bibinfo {author} {\bibfnamefont
  {Q.}~\bibnamefont {Yu}}, \bibinfo {author} {\bibfnamefont {D.}~\bibnamefont
  {Newell}}, \ and\ \bibinfo {author} {\bibfnamefont {Y.}~\bibnamefont
  {Chen}},\ }\href@noop {} {\bibfield  {journal} {\bibinfo  {journal}
  {International Journal of Modern Physics B}\ }\textbf {\bibinfo {volume}
  {27}},\ \bibinfo {pages} {1341002} (\bibinfo {year} {2013})}\BibitemShut
  {NoStop}%
\bibitem [{\citenamefont {Xia}\ \emph {et~al.}(2009)\citenamefont {Xia},
  \citenamefont {Chen}, \citenamefont {Li},\ and\ \citenamefont
  {Tao}}]{CQXIA2009}%
  \BibitemOpen
  \bibfield  {author} {\bibinfo {author} {\bibfnamefont {J.}~\bibnamefont
  {Xia}}, \bibinfo {author} {\bibfnamefont {F.}~\bibnamefont {Chen}}, \bibinfo
  {author} {\bibfnamefont {J.}~\bibnamefont {Li}}, \ and\ \bibinfo {author}
  {\bibfnamefont {N.}~\bibnamefont {Tao}},\ }\href@noop {} {\bibfield
  {journal} {\bibinfo  {journal} {Nature Nanotechnology}\ }\textbf {\bibinfo
  {volume} {4}},\ \bibinfo {pages} {505} (\bibinfo {year} {2009})}\BibitemShut
  {NoStop}%
\bibitem [{\citenamefont {Chen}\ and\ \citenamefont
  {Appenzeller}(2008)}]{CQAppenzeller2008}%
  \BibitemOpen
  \bibfield  {author} {\bibinfo {author} {\bibfnamefont {Z.}~\bibnamefont
  {Chen}}\ and\ \bibinfo {author} {\bibfnamefont {J.}~\bibnamefont
  {Appenzeller}},\ }\href@noop {} {\bibfield  {journal} {\bibinfo  {journal}
  {Proceedings of the International Electron Devices Meeting (IEDM), San
  Francisco, CA, 15–17 December 2008.}\ ,\ \bibinfo {pages} {DOI:
  10.1109/IEDM.2008.4796737}} (\bibinfo {year} {2008})}\BibitemShut {NoStop}%
\bibitem [{\citenamefont {Wenger}\ \emph {et~al.}(2009)\citenamefont {Wenger},
  \citenamefont {Lukosius}, \citenamefont {Muessig}, \citenamefont {Ruhl},
  \citenamefont {Pasko},\ and\ \citenamefont {Lohe}}]{Wenger2009}%
  \BibitemOpen
  \bibfield  {author} {\bibinfo {author} {\bibfnamefont {C.}~\bibnamefont
  {Wenger}}, \bibinfo {author} {\bibfnamefont {M.}~\bibnamefont {Lukosius}},
  \bibinfo {author} {\bibfnamefont {H.}~\bibnamefont {Muessig}}, \bibinfo
  {author} {\bibfnamefont {G.}~\bibnamefont {Ruhl}}, \bibinfo {author}
  {\bibfnamefont {S.}~\bibnamefont {Pasko}}, \ and\ \bibinfo {author}
  {\bibfnamefont {C.}~\bibnamefont {Lohe}},\ }\href@noop {} {\bibfield
  {journal} {\bibinfo  {journal} {Journal Vacuum Science and Technology B}\
  }\textbf {\bibinfo {volume} {27}},\ \bibinfo {pages} {286} (\bibinfo {year}
  {2009})}\BibitemShut {NoStop}%
\bibitem [{\citenamefont {Meric}\ \emph {et~al.}(2008)\citenamefont {Meric},
  \citenamefont {Han}, \citenamefont {Young}, \citenamefont {Ozyilmaz},
  \citenamefont {Kim},\ and\ \citenamefont {Shepard}}]{Meric2008}%
  \BibitemOpen
  \bibfield  {author} {\bibinfo {author} {\bibfnamefont {I.}~\bibnamefont
  {Meric}}, \bibinfo {author} {\bibfnamefont {M.}~\bibnamefont {Han}}, \bibinfo
  {author} {\bibfnamefont {A.}~\bibnamefont {Young}}, \bibinfo {author}
  {\bibfnamefont {B.}~\bibnamefont {Ozyilmaz}}, \bibinfo {author}
  {\bibfnamefont {P.}~\bibnamefont {Kim}}, \ and\ \bibinfo {author}
  {\bibfnamefont {K.}~\bibnamefont {Shepard}},\ }\href@noop {} {\bibfield
  {journal} {\bibinfo  {journal} {Nature Nanotechnology}\ }\textbf {\bibinfo
  {volume} {3}},\ \bibinfo {pages} {654} (\bibinfo {year} {2008})}\BibitemShut
  {NoStop}%
\bibitem [{\citenamefont {Wenger}\ \emph {et~al.}(2008)\citenamefont {Wenger},
  \citenamefont {Lupina}, \citenamefont {Lukosius}, \citenamefont {Seifarth},
  \citenamefont {Muessig}, \citenamefont {Pasko},\ and\ \citenamefont
  {Lohe}}]{Wenger2008}%
  \BibitemOpen
  \bibfield  {author} {\bibinfo {author} {\bibfnamefont {C.}~\bibnamefont
  {Wenger}}, \bibinfo {author} {\bibfnamefont {G.}~\bibnamefont {Lupina}},
  \bibinfo {author} {\bibfnamefont {M.}~\bibnamefont {Lukosius}}, \bibinfo
  {author} {\bibfnamefont {O.}~\bibnamefont {Seifarth}}, \bibinfo {author}
  {\bibfnamefont {H.-J.}\ \bibnamefont {Muessig}}, \bibinfo {author}
  {\bibfnamefont {S.}~\bibnamefont {Pasko}}, \ and\ \bibinfo {author}
  {\bibfnamefont {C.}~\bibnamefont {Lohe}},\ }\href@noop {} {\bibfield
  {journal} {\bibinfo  {journal} {Journal of Applied Physics}\ }\textbf
  {\bibinfo {volume} {103}},\ \bibinfo {pages} {104103} (\bibinfo {year}
  {2008})}\BibitemShut {NoStop}%


\end{thebibliography}
\end{document}